\documentclass[reprint,superscriptaddress,amsmath,amssymb,prx,longbibliography]{revtex4-1}

\usepackage{graphicx}
\usepackage{amsmath}
\usepackage{hyperref}
\usepackage{epsfig}
\usepackage{epstopdf}
\usepackage{amssymb}

\usepackage[ngerman, english]{babel}

\DeclareGraphicsExtensions{.pdf}
\usepackage[utf8]{inputenc}

\DeclareGraphicsExtensions{.pdf}

\usepackage{hyperref}
\hypersetup{%
	colorlinks=true,
	linkcolor=blue,
	urlcolor=blue,
	citecolor=blue
}
\begin{document}

\title{Ferromagnetic phase transition in topological crystalline insulator thin films: interplay of anomalous Hall angle and magnetic anisotropy}

\author{R. Adhikari}
\email{rajdeep.adhikari@jku.at}
\affiliation{Institut f\"ur Halbleiter-und-Festk\"orperphysik, Johannes Kepler University, Altenbergerstr. 69, A-4040 Linz, Austria}

\author{V. V. Volobuev}
\affiliation{Institut f\"ur Halbleiter-und-Festk\"orperphysik, Johannes Kepler University, Altenbergerstr. 69, A-4040 Linz, Austria}
\affiliation{International Research Centre MagTop, Institute of Physics, Polish Academy of Sciences, Aleja Lotnik\`{o}w-32/46, PL-02668 Warszawa, Poland}
\affiliation{National Technical University ''KhPI'', Kyrpychova Str. 2, 61002 Khariv, Ukraine}

\author{B. Faina}
\affiliation{Institut f\"ur Halbleiter-und-Festk\"orperphysik, Johannes Kepler University, Altenbergerstr. 69, A-4040 Linz, Austria}

\author{G. Springholz}
\affiliation{Institut f\"ur Halbleiter-und-Festk\"orperphysik, Johannes Kepler University, Altenbergerstr. 69, A-4040 Linz, Austria}

\author{A. Bonanni}
\email{alberta.bonanni@jku.at}
\affiliation{Institut f\"ur Halbleiter-und-Festk\"orperphysik, Johannes Kepler University, Altenbergerstr. 69, A-4040 Linz, Austria}

\begin{abstract}
	
In magnetic topological phases of matter, the quantum anomalous Hall (QAH) effect is an emergent phenomenon driven by ferromagnetic doping, magnetic proximity effects and strain engineering. The realization of QAH states with multiple dissipationless edge and surface conduction channels defined by a Chern number $\mathcal{C}\geq1$ was foreseen for the ferromagnetically ordered SnTe class of topological crystalline insulators (TCIs). From  magnetotransport measurements on Sn$_{1-x}$Mn$_{x}$Te ($0.00\leq{x}\leq{0.08}$)(111) epitaxial thin films grown by molecular beam epitaxy on BaF$_{2}$ substrates, hole mediated ferromagnetism is observed in samples with $x\geq0.06$ and the highest $T_\mathrm{c}\sim7.5\,\mathrm{K}$ is inferred from an anomalous Hall behavior in Sn$_{0.92}$Mn$_{0.08}$Te. The sizable anomalous Hall angle $\sim$0.3 obtained for Sn$_{0.92}$Mn$_{0.08}$Te is one of the greatest reported for magnetic topological materials. The ferromagnetic ordering with perpendicular magnetic anisotropy, complemented by the inception of anomalous Hall effect in the Sn$_{1-x}$Mn$_{x}$Te layers for a thickness commensurate with the decay length of the top and bottom surface states, points at Sn$_{1-x}$Mn$_{x}$Te as a preferential platform for the realization of QAH states in ferromagnetic TCIs.

\end{abstract}

\date{\today}
%\pacs{72.25.Dc, 72.25.Mk, 76.50.+g, 85.75.-d}

\maketitle

\section{Introduction}
 
In-depth investigations of global band topology \cite{Bansil:2016_RMP,Po:2017_Nat.Commun} have led to theoretical predictions and experimental demonstrations of new states of topologically protected quantum states in condensed matter systems \cite{Wang:2017_Nat.Mater.,Wen:2017_RMP}. This new family of materials hosts a wide range of quantum quasiparticles like \textit{e.g} Dirac, Weyl and Majorana fermions \cite{Pal:2011_AJP}. Material classes such as topological insulators (TI) \cite{Hasan:2010_RMP,Kane:2005_PRL}, topological crystalline insulators (TCI) \cite{Fu:2011_PRL,Ando:2013_JPSJ,Ando:2015_Ann.Rev.}, topological superconductors (TSC) \cite{Ando:2015_Ann.Rev., Sato:2017_Rep.Prog.Phys}, Weyl semimetals (WSM) \cite{Armitage:2018_RMP,Soluyanov:2015_Nature} and Dirac semimetals (DSM) \cite{Armitage:2018_RMP,Bernevig:2018_JPSJ,Wehling:2014_Adv.Phys.} defined by their band topology are expected to play a major role in quantum technology. The quantization of the Hall conductance observed in a 2-dimensional electron gas (2DEG) \cite{Klitzing:1980_PRL} challenged the time honoured Landau paradigm \cite{Ginzburg:1950_ZETF} for ordered states of matter classified according to spontaneously broken symmetries \cite{Wen:2017_RMP}. An external magnetic field perpendicular to a 2DEG forces the electrons to undergo circular motion in quantized orbits, defined by the corresponding Landau levels (LLs), leading to insulating bulk and conducting edge states. These edge states differ from trivial states of matter particularly in their robustness against impurities. The Hall conductance in a 2DEG is quantized and defined as $\sigma_{\mathrm{xy}}=\left(\frac{ne^{2}}{h}\right)$, where $n$ is an integer, $e$ the electronic charge, $h$ the Planck's constant and $R_{\mathrm{K}}$=$\frac{h}{e^{2}}$ the von Klitzing constant. The quantization of the Hall conductance has been found to be accurate to within 1 part in 10$^{9}$ of an $1~\Omega$ resistor \cite{Poirier:2011_CRP}. 

In the seminal work of Thouless and co-workers \cite{Thouless:1982_PRL}, the quantum Hall states in a 2DEG are identified to be of topological origin and are characterized by robust edge states, rendering the difference between a quantum Hall state and an ordinary insulator state a matter of topology. Topological quantum matter can be classified as$\colon$ $\left( i\right)$ topologically protected  \cite{Thouless:1982_PRL,Hasan:2010_RMP} or $\left( ii\right)$  topologically ordered \cite{Wen:1995_Adv.Phys.}. The topologically protected states are characterized by an unique ground state preserved by a topological invariant, by ordinary electron excitations or by short-range quantum entanglement \cite{Hasan:2010_RMP, Ando:2013_JPSJ}. On the other hand, topologically ordered matter has ground state degeneracies on higher genus manifolds, fractionalized excitations and long-range quantum entanglement. While the integer quantum Hall (IQH) states are topologically protected phases, the fractional quantum Hall (FQH) states belong to the class of topologically ordered ones. A topological state like the IQH has response function characterized by a topological invariant $n\in\mathbb{Z}$, defined as a quantity that remains unchanged under continuous deformation or homeomorphisms \cite{Nakahara:Book}. The integer $n$ appearing in the mentioned definition of quantized Hall conductance $\sigma_{\mathrm{xy}}$ corresponds to the number of Landau levels under the Fermi level $E_{\mathrm{F}}$. Each LL contributes one $\frac{e^2}{h}$ quantum to the total Hall conductance. The Thouless-Kohmoto-Nightingale-Nijis (TKKN) invariant $n$ \cite{Kohmoto:1985_Ann.Phys.,Thouless:1982_PRL,Hasan:2010_RMP,Ando:2013_JPSJ} is also the first Chern number $\mathcal{C}$. 

The integer $\mathcal{C}$ has its origin in topology and according to the Gauss-Bonnet theorem, the integral of the curvature of a mathematical manifold over a geometric surface is a topological invariant \cite{Hasan:2010_RMP,Nakahara:Book}. In condensed matter systems, the integral of the Berry flux $\mathcal{F_\mathrm{b}}$ over closed surfaces like a sphere or a torus, is the topological invariant $\mathcal{C}$. The quantity $\mathcal{F_\mathrm{b}}$ has its origin in the Berry phase $\mathcal{A_\mathrm{b}}$, which is a gauge-invariant, geometric phase acquired by a system under a cyclic and adiabatic process \cite{Xiao:2010_RMP}, and $\mathcal{F_\mathrm{b}}=\nabla\times\mathcal{A_\mathrm{b}}$. The topological invariant $\mathcal{C}$ is related to $\mathcal{A_\mathrm{b}}$ as:

\begin{equation}
\mathcal{C}=\frac{1}{2\pi}\int{\mathrm{d}^{2}\textbf{k}~\mathcal{F_\mathrm{b}}}
\end{equation}
\noindent
where $\textbf{k}$ is the wave vector. The dependence of  $\sigma_\mathrm{xy}$ on $\mathcal{C}$ explains the robust quantization of the QHE.

%%%%%%%%%%%%%%%%%%%%%%%%%%%%%%%%%%%%%%%%%%%%%%%%%%%%%%%%%%%%%%%%%%%%%
\begin{figure*}[htbp]
	\centering
	\includegraphics[scale=0.5]{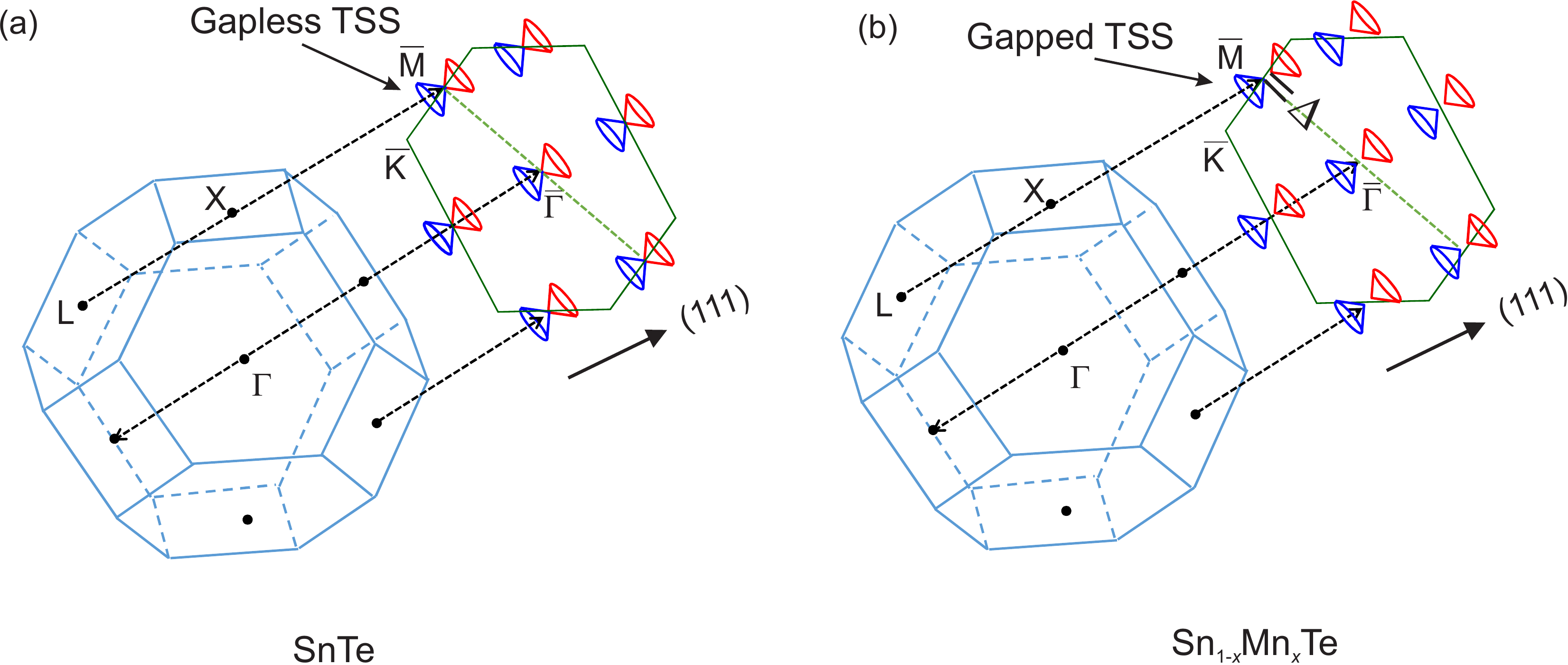}
	%\vspace{3 cm}
	\caption{\label{fig:Dirac} (a) Brillouin zone of undoped SnTe (111) grown on a BaF$_{2}$ substrate with gapless Dirac cones at the TRIM of the BZ. (b) Brillouin zone of the Sn$_{1-x}$Mn$_{x}$Te with gapped Dirac cones at the TRIM due to a breaking of the $\mathcal{T}$ symmetry. $\Delta$: Energy gap of the Dirac cone.}	
	\label{fig:fig1}
\end{figure*}
%%%%%%%%%%%%%%%%%%%%%%%%%%%%%%%%%%%%%%%%%%%%%%%%%%%%%%%%%%%%%%%%%%%%%

In the case of the IQH effect, since $\sigma_{\mathrm{xy}}$ is odd under time reversal symmetry $\mathcal{T}$, the non-trivial topological states only occur upon $\mathcal{T}$ symmetry breaking \cite{Kohmoto:1985_Ann.Phys.}. However, the presence of spin-orbit coupling (SOC) allows a distinct class of topological and non-trivial insulators that preserve the $\mathcal{T}$ symmetry \cite{Kane:2005_PRL, Hasan:2010_RMP, Qi:2010_RMP}. The $\mathcal{T}$ invariant Bloch Hamiltonians and their equivalent classes can be smoothly transformed without closing the energy gap and generate the $\mathcal{T}$ invariant topological insulators \cite{Hasan:2010_RMP, Qi:2010_RMP}, characterized by the topological invariant $\mathbb{Z}_{2}$ \cite{Bernevig:2006_Science,Bernevig:2006_PRL}. The strong SOC in CdTe/HgTe quantum well structures leads to band inversion and to the non-trivial phase of quantum spin Hall (QSH) insulator \cite{Bernevig:2006_Science,Bernevig:2006_PRL}, which was the first observed 2D TI  \cite{Koenig:2007_Science}. The concept of symmetry protected topological (SPT) phase of matter \cite{Senthil:2015_Ann.Rev.} was predicted \cite{Fu:2007_PRL, Zhang:2009_Nat.Phys.} and realized in 3-dimensional (3D) systems of Bi$_{1-x}$Sb$_{x}$ \cite{Hsieh:2008_Nature}, Bi$_{2}$Se$_{3}$ and Bi$_{2}$Te$_{3}$ \cite{Xia:2009_Nat.Phys.}.

\section{Background}

The TCIs represent topological phases protected by crystal symmetries \cite{Fu:2011_PRL,Hsieh:2012_Nat.Commun.,Ando:2015_Ann.Rev.,Ando:2013_JPSJ,Khalaf:2018_PRX,Safaei_:2015_NJP}, such as reflection symmetry with respect to a crystal plane, \textit{i.e.} mirror symmetry $\mathcal{M}$. The presence of $\mathcal{M}$ symmetry has fundamental implications in topology, since it yields two independent topological invariants, namely the total Chern number $\mathcal{C}_{\mathrm{t}}$ and the mirror Chern number $\mathcal{C}_{\mathcal{M}}$. Therefore, even in the absence of $\mathcal{C}_{\mathrm{t}}$, a robust topological phase in TCI can still be defined by $\mathcal{C}_{\mathcal{M}}$. The surfaces of TCIs host gapless states that depend on the surface orientations of the TCI crystal. In a nutshell, the TCIs are the counterpart of TIs in material systems lacking SOC. The IV-VI low bandgap semiconductor SnTe was the first material system theoretically predicted to host topological surface states protected by $\mathcal{M}$ symmetry \cite{Hsieh:2012_Nat.Commun.}, as then proven by angle resolved photoemission spectroscopy (ARPES) experiments \cite{Tanaka:2012_Nat.Phys.}. Bulk SnTe was shown to possess robust surface states with even number of Dirac cones at the crystal planes $\left(100\right)$, $\left(110\right)$ and $\left(111\right)$. A crystal symmetry protected topological behavior was also observed in Pb$_{1-x}$Sn$_x$Se \cite{Dziawa:2012_Nat.Mater} and Pb$_{1-x}$Sn$_x$Te \cite{Xu:2012_Nat.Commun.,Assaf:2016_Sci.Rep,Volobuev:2017_Adv.Mater.}. The topological character of SnTe is due to a non-zero $\mathcal{C_{\mathcal{M}}}$ with $|\mathcal{C}_{\mathcal{M}}|=2$. 
The TCI-like behavior of the SnTe class of materials including Pb$_{1-x}$Sn$_{x}$Te, SnSe, and Pb$_{1-x}$Sn$_{x}$Se led to an extensive research effort to understand their electronic and topological properties  \cite{Ando:2015_Ann.Rev.,Tanaka:2012_Nat.Phys.,Dziawa:2012_Nat.Mater,Xu:2012_Nat.Commun.}. The IV-VI compounds have been widely studied as thermoelectric materials, as photodetectors and as dilute magnetic semiconductors \cite{Dimmock:1966_PRL,Story:1990_SST,Chi:2016_APL,Nadolny:2002_JMMM,Eggenkamp:1994_JAP}. The interest in the topological behavior of undoped and magnetically doped IV-VI compounds, both in the bulk and thin layer arrangement, has significantly increased recently \cite{Wei:2018_PRB,Nadolny:2002_JMMM,Dybko:2017_PRB,Eggenkamp:1994_JAP,Chi:2016_APL,Reja:2017_PRB,Wang:2018_PRB,Volobuev:2017_Adv.Mater.,Assaf:2016_Sci.Rep}. By means of ARPES, it was demonstrated, that the four Dirac cones of the TCIs are spin non-degenerate and helically spin-polarized \cite{Xu:2012_Nat.Commun.,Ando:2013_JPSJ,Ando:2015_Ann.Rev.}. These compounds crystallize in a cubic rocksalt structure and can be cleaved along the $\left(100\right)$ and the $\left(111\right)$ planes. Spectroscopic studies on samples cleaved or grown along the $\left(100\right)$ plane show a double Dirac cone structure near the $\overline{X}$ point of the Brillouin zone (BZ). These Dirac points of the $\left(100\right)$ surface are not located at the time-reversal-invariant-momenta (TRIMs). The $\mathcal{M}$ symmetry-constrained hybridization of the two Dirac cones leads to Lifshitz transitions in $\left(100\right)$-oriented TCIs \cite{Ando:2015_Ann.Rev.}. Furthermore, it has been predicted by \textit{ab initio} calculations \cite{Liu:2013_PRB} and experimentally confirmed \cite{Taskin:2014_PRB}, that all four Dirac points are located at the TRIMS for the $\left(111\right)$ surface of the SnTe compounds. The isotropic Dirac surface state is located at the $\overline{\Gamma}$ point, while three anisotropic Dirac states are centered at the $\overline{\mathrm{M}}$ point \cite{Liu:2015_PRB,Wang:2018_PRB,Li:2016_PRL}. The difference in energy and Fermi velocity of charge carriers for the different Dirac points at $\overline{\Gamma}$ and $\overline{\mathrm{M}}$, together with the $\mathcal{T}$ symmetry, induces valley-dependent surface transport effects in the SnTe class of materials. Recently, it was demonstrated that the topological properties of SnTe can be tuned by applying stress \cite{Schreyeck:2019_PRM} or $via$ thickness modulation \cite{Liu:2013_Nat.Mater.}. Moreover, it was shown both by theoretical calculations \cite{Brzezicki:2018_arxiv} and by employing magnetotransport experiments \cite{Mazur:2018_arxiv}, that Majorana-like excitations can be stabilized at the monoatomic surface steps of TCI systems like SnTe and Pb$_{1-x}$Sn$_{x}$Te . 

The counter-propagating dissipationless helical edge channels with opposite spins that are robust against non-magnetic disorder lead to the realization of QSH states in SPT phases of matter. Magnetic doping of SPTs -- such as TIs and TCIs -- drives the non-trivial surface states massless Dirac fermions (DF) into a massive DF state \cite{Kou:2015_SSC,Liu:2016_Ann.Rev.}. Due to the relation of the intrinsic anomalous Hall effect (AHE) and of the QHE with the geometric Berry phase, a ferromagnetic (FM) insulator with nonzero $\mathcal{C}$ gives rise to quantum anomalous Hall (QAH) states \cite{Nagaosa:2010_RMP,Liu:2016_Ann.Rev.}, due to the quantization of $\sigma_\mathrm{xy}$ in the absence of an external field $\mu_{0}H$. A similar dissipationless edge conduction without the requirement of $\mu_{0}H$ was theoretically proposed \cite{Haldane:1988_PRL} for a graphene monolayer \cite{Novoselov:2004_Science}. Although the Haldane model \cite{Haldane:1988_PRL} points at the theoretical possibility of realizing QAH states in a spin degenerate Dirac semimetal such as graphene, this topological effect was first observed in a QSH insulator, specifically in Mn-doped CdTe/HgTe/CdTe quantum well (QW) heterostructures \cite{Zhang:2014_PRL}. However, due to the paramagnetic behavior of Mn spins in HgTe QWs, an external $\mu_{0}H$ is still required to polarize them in order to observe the QAH effect.

The non-trivial topology of the QAH states arises from the coherent coupling of the electronic spin and charge states with the magnetic order of the magnetically doped system. In 3D TIs such as Bi$_{2}$(Se,Te)$_{3}$, each surface hosts gapless 2D Dirac spin spilt states and is the workbench to study QAH states in SPT phases. Magnetic doping of 3D TIs leads to an opening of the gap at the DPs, due to the out-of-plane Zeeman fields giving rise to a gapped Dirac cone  \cite{Tokura:2019_Nat.Rev.Phys.,Zhang:2013_Science,Yu:2010_Science,Chang:2013_Science,Tokura:2019_Nat.Rev.Phys.} which contributes to $\sigma_\mathrm{xy}=\pm\frac{e^{2}}{2h}$. Hence, in a 3D TI thin film, the resulting $\mathcal{C}=\pm1$ is due to the contributions of both the top and bottom surfaces to $\sigma_\mathrm{xy}$. A QAH state with $\left|\mathcal{C}\right|\textgreater 1$ is expected in systems with additional (to time reversal) symmetries like $e.g.$ the TCIs \cite{Tokura:2019_Nat.Rev.Phys.,Fang:2014_PRL,Wang:2018_PRB}. The exchange coupling of TCI surface Dirac fermions and magnetic dopants results in a $\mathcal{T}$ symmetry breaking and the out-of-plane magnetization opens up the Zeeman gaps of the Dirac cones, essential for the realization of high-$\mathcal{C}$ QAH states in magnetic TCIs. By driving a TCI into an ordered FM state with perpendicular magnetic anisotropy (PMA), QAH states with $\mathcal{C}$\,=$\pm$1 can be induced in magnetically doped TCI thin films \cite{Ando:2015_Ann.Rev.,Fang:2014_PRL}. A high-$\mathcal{C}$ QAH state with $\mathcal{C}$\,=$\pm$4 emerges in a magnetically doped TCI thin film as a result of the interplay between crystalline symmetry, induced Zeeman field from the FM ordering, strain engineering, and an adjusted film thickness enabling the top and bottom surface states to interact. In comparison to the single chiral edge channel of the $\mathcal{C}$\,=$\pm$1 QAH states of magnetically doped TI, high-$\mathcal{C}$ QAH systems like magnetic TCI are expected to host multiple dissipationless topologically protected chiral edge conduction channels. The $\mathcal{M}$ symmetry-protected topological nature of TCI and the presence of an even number of gapless Dirac surface states leads to topological phenomena like spin-filtered multiple edge states, QSH effect and high-$\mathcal{C}$ QAH states that are envisioned to be particularly relevant for quantum technology and quantum metrology \cite{Fox:2018_PRB}.

%%%%%%%%%%%%%%%%%%%%%%%%%%%%%%%%%%%%%%%%%%%%%%%%%%%%%%%%%%%%%%%%%%%%%
\begin{figure*}[htbp]
	\centering
	\includegraphics[scale=0.35]{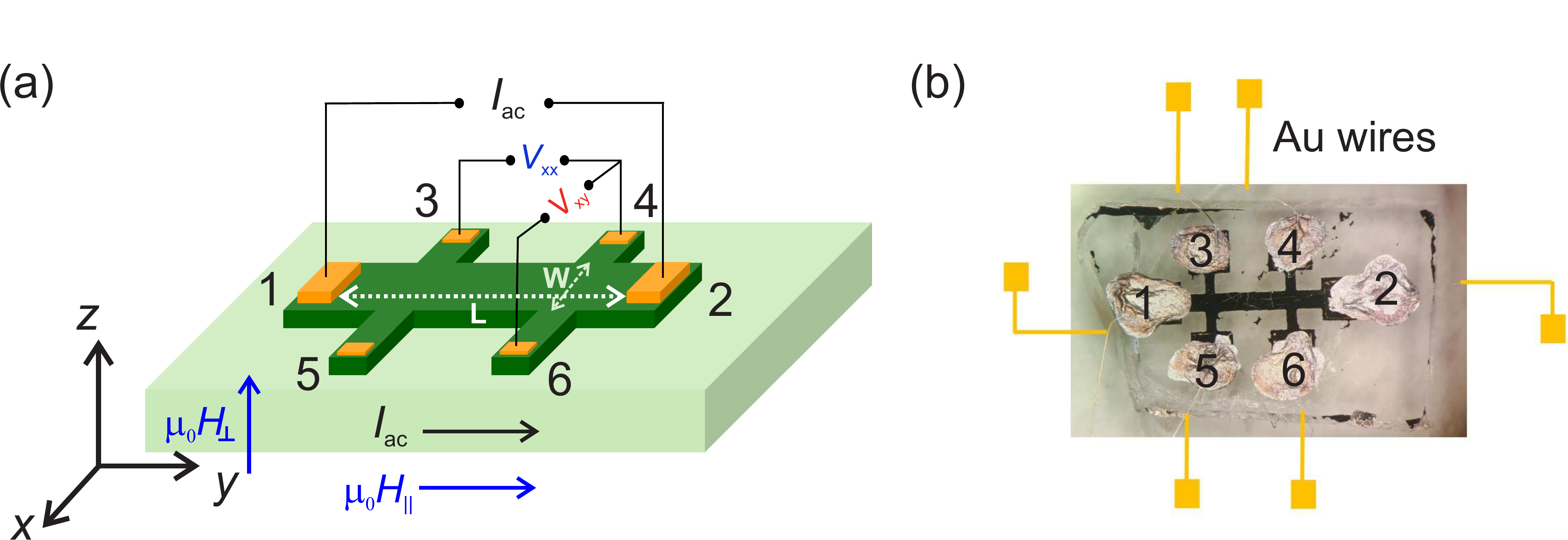}
	%\vspace{3 cm}
	\caption{\label{fig:Device} Hall bar devices used for the magnetotransport measurements: (a) sketch; (b) photograph. $V_{\mathrm{xx}}$: voltage due to longitudinal resistance; $V_{\mathrm{xy}}$: Hall voltage; $I_{ac}$: applied in-plane current; $H_{\parallel}$: parallel magnetic field and and $H_{\perp}$: perpendicular magnetic field. x-y-z defines the geometry of the sample w.r.t. the applied current and magnetic fields. }
	
	\label{fig:fig2}
\end{figure*}
%%%%%%%%%%%%%%%%%%%%%%%%%%%%%%%%%%%%%%%%%%%%%%%%%%%%%%%%%%%%%%%%%%%%% 
 
Among the magnetic impurities in TCIs, the most widely studied are the transition metals Mn and Cr \cite{Tokura:2019_Nat.Rev.Phys.,Wang:2018_PRB,Nadolny:2002_JMMM,Eggenkamp:1994_JAP,Chi:2016_APL,Story:1990_SST}. Rich magnetic phases exhibiting paramagnetic, spin glass (SG), FM and re-entrant spin glass (RSG) behaviour have been reported for bulk and thin Sn$_{1-x}$Mn$_{x}$Te films \cite{Wang:2018_PRB,Nadolny:2002_JMMM,Eggenkamp:1994_JAP,Eggenkamp:1995_PRB,Story:1993_PRB}. An interplay between Mn- and hole-concentration in Sn$_{1-x}$Mn$_{x}$Te determines the magnetic behavior of the system leading to Rudermann-Kittel-Kasuya-Yoshida (RKKY) carrier mediated ferromagnetism \cite{Eggenkamp:1995_PRB}. Along with the RKKY interaction, the magnetic ordering in Sn$_{1-x}$Mn$_{x}$Te has been shown to originate from other mechanisms like superexchange and direct exchange, depending on the distributions of Mn ions in the SnTe lattice \cite{Eggenkamp:1995_PRB,Eggenkamp:1994_JAP,Wang:2018_PRB,Story:1993_PRB}. In-depth studies of the magnetotransport properties of thin epitaxial Sn$_{1-x}$Mn$_{x}$Te layers to understand the correlation between AHE, the anomalous Hall angle $\theta_{\mathrm{AH}}$, PMA, the magnetic dopant concentration $x$ and the film thickness are essential for the realization of high-$\mathcal{C}$ QAH states.\\ 

Here, these correlations are investigated by magnetotransport in Sn$_{1-x}$Mn$_{x}$Te epitaxial layers grown by molecular beam epitaxy (MBE) on insulating BaF$_{2}$ (111) substrates. A critical film thickness comparable with the decay length of the top and bottom surface states, leads to  hybridization of the total $\sigma_\mathrm{xy}$, a fundamental requirement for a quantum phase transition and for the realization of high-$\mathcal{C}$ QAH states in the considered Mn doped TCI layers. A sketch of the BZ of undoped SnTe (111) with the TRIM points is shown in Fig.~\ref{fig:fig1}(a). Doping of SnTe with Mn breaks the $\mathcal{T}$ symmetry and the Zeeman field in the electronic band structure of Sn$_{1-x}$Mn$_{x}$Te leads to gapped Dirac cones at the TRIM points -- one at the $\overline{\Gamma}$ point and three at the $\overline{\mathrm{M}}$. If the chemical potential is adjusted in all Zeeman gaps at the four TRIM points simultaneously, a QAH state with high Chern number can be realized. The BZ of the Sn$_{1-x}$Mn$_{x}$Te (111) with the gapped Dirac cones at the $\overline{\Gamma}$ and $\overline{\mathrm{M}}$ points is shown in Fig.~\ref{fig:fig1}(b). The realization of high-$\mathcal{C}$ magnetic TCIs requires a system satisfying the following conditions:\\*
(i) presence a FM ordering;\\
(ii) emergence of AHE in the FM ordered state;\\
(iii) anisotropy in the magnetic properties manifested by PMA;\\
(iv) adjusted thickness commensurate with the decay length of the topological top and bottom surface states. 

A series of 30 nm thin Sn$_{1-x}$Mn$_{x}$Te layers have been grown directly on insulating BaF$_{2}$ (111) substrates without buffer between substrate and overlayer, in contrast to previous experimental reports \cite{Wang:2018_PRB,Chi:2016_APL}, where the magnetically doped TCI layer was grown on a buffer of undoped SnTe. The absence of the buffer layer in the structures studied here guarantees that the electrical transport takes place only through the Sn$_{1-x}$Mn$_{x}$Te channel. As discussed earlier, the choice of a $\left(111\right)$ growth direction ensures that the topological character of the pristine SnTe system is due to both $\mathcal{M}$ symmetry and $\mathcal{T}$ invariance \cite{Liu:2015_PRB,Wang:2018_PRB,Liu:2013_Nat.Mater.}. A film thickness of 30 nm is selected, since it is on one hand compatible with the decay length of the topological surface states and on the other hand it ensures high crystallinity. Magnetotransport measurements are performed on the epitaxial Sn$_{1-x}$Mn$_{x}$Te layers and yield $\sigma_{\mathrm{xx}}$ and $\sigma_{\mathrm{xy}}$ as a function of $T$ and $\mu_{0}H$. By varying the concentration of Mn in the SnTe lattice, the onset of a stable hole-mediated FM interaction in the epitaxial Sn$_{1-x}$Mn$_{x}$Te thin layers is found for $x\geq0.06$. The FM ordering is assessed from the observed hysteretic behavior of $\sigma_{\mathrm{xx}}$ and $\sigma_{\mathrm{xy}}$ and the Curie temperature $T_{\mathrm{c}}$ scales as a function of $x$. The interplay between $\theta_{\mathrm{AH}}$, magnetic anisotropy, $T_{\mathrm{c}}$, carrier concentration and $x$ triggers a FM phase transition in the thin epitaxial Sn$_{1-x}$Mn$_{x}$Te TCI layers.

\section{Methods}  
   
\subsection{Sample growth and Hall bar fabrication} 

%%%%%%%%%%%%%%%%%%%%%%%%%%%%%%%%%%%%%%%%%%%%%%%%%%%%%%%%%%%%%%%%%%%%%
\begin{figure*}[htbp]
	\centering
	\includegraphics[scale=2.5]{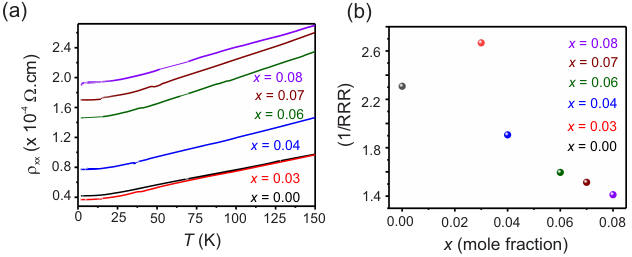}
	%\vspace{3 cm}
	\caption{\label{fig:Resistivity} (a) $\rho_{xx}$ as a function of $T$ for Sn$_{1-x}$Mn$_{x}$Te layers. (b) Inverse of residual resistivity ratio as a function of $x$.}	
	\label{fig:fig3}
\end{figure*}
%%%%%%%%%%%%%%%%%%%%%%%%%%%%%%%%%%%%%%%%%%%%%%%%%%%%%%%%%%%%%%%%%%%%% 
 
The Sn$_{1-x}$Mn$_{x}$Te epitaxial layers are grown on BaF$_{2}$ (111) substrates in a GEN-II Veeco MBE system using stoichiometric SnTe, Te and Mn sources. The investigated samples series consists of 30 nm thick highly crystalline Sn$_{1-x}$Mn$_{x}$Te films with Mn concentrations $x=0.03$, $0.04$, $0.06$, $0.07$, $0.08$ and of a reference SnTe film. As reported in Refs.\cite{Chi:2016_APL,Nadolny:2002_JMMM}, the electronic properties of SnTe depend critically on the density of Sn vacancies $V_\mathrm{Sn}$ in the lattice. Each $V_\mathrm{Sn}$ acts as double acceptors, leading to $p$--type conductivity in both bulk and thin layers of pristine SnTe. The Sn vacancies and thus the hole concentration in SnTe can be controlled and tuned by adjusting the Te flux during deposition. The desired Mn concentrations have been achieved by regulating the Mn flux calibrated according to well-established growth protocols \cite{Volobuev:2017_Adv.Mater.,Story:1990_SST}. The actual Mn concentrations have been confirmed $via$ high-resolution x-ray diffraction (HRXRD) measurements.  In the present study, the growth of the Sn$_{1-x}$Mn$_{x}$Te thin films is carried out in excess of Te flux in order to control the stoichiometry of the crystal. The film thickness is determined on-line with a quartz microbalance \cite{Volobuev:2017_Adv.Mater.} and the epitaxial process is monitored by means of \textit{in situ} reflection high-energy electron diffraction (RHEED). A protocol of surface and structural characterization, including atomic force microscopy (AFM) and HRXRD provides control on the quality of the layers and confirms the absence of Mn-precipitation, ruling out the formation of secondary phases.  

The Sn$_{1-x}$Mn$_{x}$Te films are patterned as symmetric Hall bars \textit{via}  photolithography. The samples are first spin coated with S1805 positive photoresist and then soft baked at 90$^{\circ}$C. A S\"uss Mask aligner photolithography system has been employed to expose the S1805 coated samples under an ultra-violet mercury lamp through a Hall bar window mask. The exposed samples are developed using a TMAH developer and 1-2-2-1 Hall bar structures are chemically etched using a bromomethanol solution. The etched Hall device mesas have a channel length $L=1700\,\mu$m and a width $W=450\,\mu$m.  After wet chemical etching, the photoresist is stripped off by rinsing the samples in acetone at room temperature. The samples are then cleaned in de-ionized water and dried under high purity compressed $\mathrm{N}_{2}$ flow. 

The Ohmic contacts to the Sn$_{1-x}$Mn$_{x}$Te films consist of a 10 nm Ti layer followed by 50 nm Au, deposited by e-beam evaporation in a base pressure of $\left( 1 \times 10^{-6}\right)$ mbar in a Belzar e-beam evaporation system. The Ohmic contact pads to the Hall bars are also fabricated by photolithography and care is taken that the Sn$_{1-x}$Mn$_{x}$Te films do not undergo oxidation prior to deposition of the Ti/Au metal contacts. The linear Ohmic characteristics of the Ti/Au contacts to the Sn$_{1-x}$Mn$_{x}$Te layers are tested with a Keithley 4200 SCS system.

\subsection{Magnetotransport}

%%%%%%%%%%%%%%%%%%%%%%%%%%%%%%%%%%%%%%%%%%%%%%%%%%%%%%%%%%%%%%%%%%%%%
\begin{figure*}[htbp]
	\centering
	\includegraphics[scale=2.0]{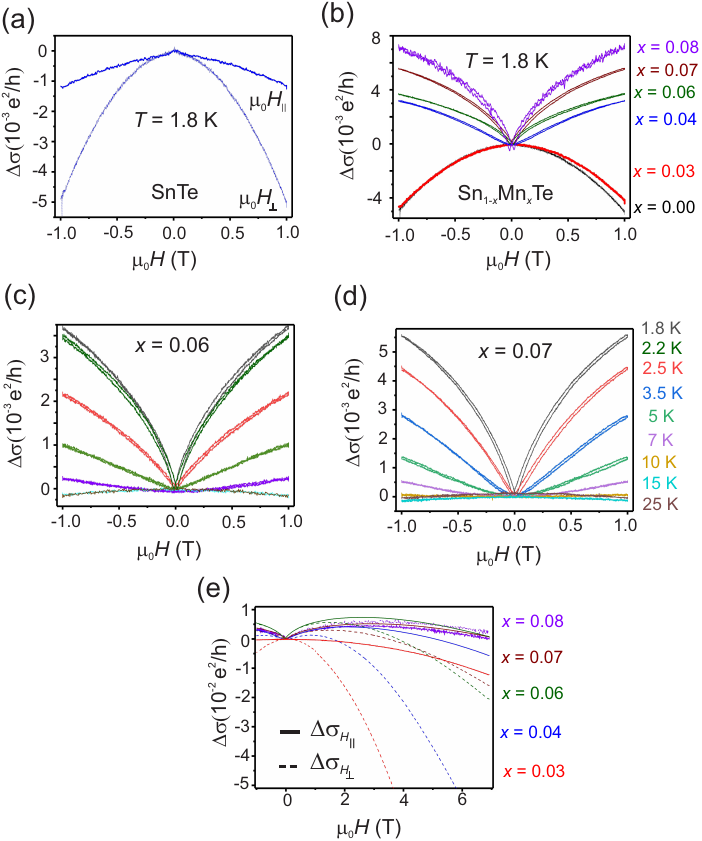}
	%\vspace{3 cm}
	\caption{\label{fig:MC} (a) Magnetoconductance of SnTe film measured for $H_{\parallel}$ and $H_{\perp}$ geometries at $T=1.8\,\mathrm{K}$. (b) Magnetoconductance of the Sn$_{1-x}$Mn$_{x}$Te thin films at $T = 1.8\,\mathrm{K}$. Magnetoconductance as a function of $T$ for Sn$_{1-x}$Mn$_{x}$Te thin films with (c) $x=0.06$, and (d) $x=0.07$ recorded for the in-plane $H_{\parallel}$ geometry. (e) Magnetic anisotropy of Sn$_{1-x}$Mn$_{x}$Te at $T=1.8\,\mathrm{K}$, with $\Delta\sigma$ measured for $\mu_{0}H_\parallel$ and $\mu_{0}H_\perp$.}	
	\label{fig:fig4}
\end{figure*}
%%%%%%%%%%%%%%%%%%%%%%%%%%%%%%%%%%%%%%%%%%%%%%%%%%%%%%%%%%%%%%%%%%%%%

A schematic of the measured Hall devices and the photograph of an actual device are shown in Figs.~\ref{fig:fig2}(a) and \ref{fig:fig2}(b), respectively. The quantities $I_{\mathrm{ac}}$, $V_{\mathrm{xx}}$ and $V_{\mathrm{xy}}$ refer to the applied in-plane current, longitudinal and Hall voltages, while $H_{\parallel}$ and $H_{\perp}$ indicate the parallel and perpendicular applied magnetic fields. Electrically conducting Au wires with a diameter of 25 $\mu$m are bonded with highly conducting Ag epoxy to the Ti/Au Ohmic contacts, as shown in Fig.~\ref{fig:fig2}(b).  
 
Low $T$-high $\mu_{0}H$ magnetotransport measurements are carried out in a Janis Super Variable Temperature 7TM-SVM cryostat equipped with a 7\,T superconducting magnet. A lock-in amplifier (LIA) \textit{ac} technique is employed for measuring the magnetotransport properties of the Sn$_{1-x}$Mn$_{x}$Te thin films. The current is sourced from a Stanford Research SR830 LIA, while the longitudinal voltage $V_{\mathrm{xx}}$ and the Hall voltage $V_{\mathrm{xy}}$ are measured in a phase locked mode. The lock-in expand function is employed to enhance the sensitivity of the LIA. As evidenced in Fig.~\ref{fig:fig2}(a), $I_{\mathrm{ac}}$ is sourced between the leads 1--2, while $V_{\mathrm{xx}}$ and $V_{\mathrm{xy}}$ are measured across the leads 3--4 and 4--6, respectively. All measurements have been performed at a frequency of 127\,Hz. An input current of 10\,$\mu$A is chosen for all the measurements, unless otherwise mentioned. The low input current is essential to avoid thermal drift due to Joule heating of the samples. The magnetic fields have been varied between -1\,T and +1\,T for the Hall measurements, while the magnetoconductance (MC) is measured by sweeping $\mu_{0}H$ between -1\,T and +7\,T. All experimental set-ups including the LIA, the magnet power supply and the temperature controller are regulated by means of an indigenously developed software.

\section{Results and Discussions}

 \subsection{Longitudinal resistivity and magnetoconductance} 
 
 The electronic properties of topological systems like TIs and TCIs are determined by both the surface and the bulk states. The penetration depth of the surface states is generally few tens of nm \cite{Zhang:2018_Sci.Bull.} and the transport is dominated by the bulk states. The surface and bulk contributions to the electronic behavior of the topologically protected states can be directly probed by combining $T$-, $\mu_{0}H$- and electrical field-dependent magnetotransport measurements.  A direct method to distinguish between the contributions of the surface and of the bulk states consists in probing the dimensionality of the Fermi surface (FS) by exploring the quantum oscillations of the longitudinal resistivity $\rho_{\mathrm{xx}}$ or of the conductance $\sigma_{\mathrm{xx}}$ as a function of the applied $\mu_{0}H$. For magnetically doped SPTs, the analysis of the $T$- and $\mu_{0}H$-dependent $\sigma_{\mathrm{xx}}$, $\sigma_\mathrm{xy}$, and of the anomalous Hall conductance $\sigma_\mathrm{xy}^{\mathrm{AH}}$ provides quantitative signatures of topology driven phase transitions \cite{Ando:2013_JPSJ,Ando:2015_Ann.Rev.,Wang:2018_PRB,Fu:2011_PRL,Fang:2014_PRL}. The evolution of $\rho_{\mathrm{xx}}$, measured as a function of $T$ in the interval $1.5\,\mathrm{K}<T<150\,\mathrm{K}$, for the Sn$_{1-x}$Mn$_{x}$Te thin films is reported in Fig.~\ref{fig:fig3}(a). A metallic behavior is observed for all Sn$_{1-x}$Mn$_{x}$Te layers, in agreement with previous reports on Sn$_{1-x}$Mn$_{x}$Te bulk crystals and thick films \cite{Chi:2016_APL,Dybko:2017_PRB,Nadolny:2002_JMMM}. The absence of an increment of $\rho_{\mathrm{xx}}$ for  $T\textless10~\mathrm{K}$ indicates that the electronic ground state of the Mn doped TCI system SnTe is not insulating, in contrast to $e.g.$ the case of Cr-doped Bi$_{2}$Se$_{3}$ \cite{Liu:2012_PRL} and SnTe \cite{Wang:2018_PRB}. In particular, for thin Sn$_{0.88}$Cr$_{0.12}$Te films, a combination of electron-electron interaction and topological delocalization leads to the insulating ground state \cite{Wang:2018_PRB}. The $\rho_{\mathrm{xx}}$\,$vs.$\,$T$ plots reported in Fig.~\ref{fig:fig3}(a) show a monotonic increase in the resistivity with Mn doping. For $T\textless$75\,K, however, Sn$_{0.97}$Mn$_{0.03}$Te is less resistive than SnTe, as a result of the increment the carriers (holes) concentration in Sn$_{0.97}$Mn$_{0.03}$Te, compared to the pristine SnTe. In order to qualitatively describe the dependence of disorder on the Mn content in Sn$_{1-x}$Mn$_{x}$Te, a modified residual resistivity ratio ($\mathrm{RRR}$) is defined as $\mathrm{RRR}=\rho_{\mathrm{xx}}(\mathrm{150~K})/\rho_{\mathrm{xx}}(\mathrm{1.8~K})$. The quantity  $\Big(\frac{1}{\mathrm{RRR}}\Big)$ is expected to decrease with increasing Mn concentration in the SnTe lattice. The plot of $\big(\frac{1}{\mathrm{RRR}}\big)$ as a function of $T$ reported in Fig.~\ref{fig:fig3}(b), shows a reduction in $\Big(\frac{1}{\mathrm{RRR}}\Big)$ with $x$ and points at an enhancement of disorder in the Sn$_{1-x}$Mn$_{x}$Te lattice with increasing Mn content, as confirmed by the HRXRD measurements.\\ 
 
As previously stated, in-depth information on the electronic behavior is gained by studying the magnetotransport properties as a functions of $T$ and $\mu_{0}H$ and from the analysis of MC, $\Delta\sigma$, and of $\sigma_{\mathrm{xy}}$. The $\Delta\sigma$ of the studied films is given by: 
 
 \begin{equation}
 \Delta \sigma =\Big[\sigma_{\mathrm{xx}}(H)-\sigma_{\mathrm{xx}}(0)\Big]\Big(\frac{\mathrm{e^2}}{h}\Big)
 \end{equation}
\noindent 
where $\sigma_{\mathrm{xx}}(H)$ and $\sigma_{\mathrm{xx}}(0)$ represent the magnetoconductance of the system under an applied field $\mu_{0}H$ and in zero field, respectively. The values of $\sigma_{\mathrm{xx}}(H)$ are estimated from the measured resistance and from the Hall bar geometry according to the relation:

\begin{equation}
\sigma_{\mathrm{xx}}=\Big(\frac{L}{W}\Big)\left[ \mathrm{R}_{xx}\left(\frac{e^{2}}{h}\right)\right ]
\end{equation}
\noindent
where $L$ and $W$  are the length and width of the Hall bar --as reported in Fig.~\ref{fig:fig2}(a)-- while $\Big(\frac{\mathrm{e^2}}{h}\Big)$ is the inverse of $R_\mathrm{K}$. The external magnetic field has been applied in two distinct orientations, namely: (i) along the $x$--axis in the $x$-$y$ plane, parallel to $I_\mathrm{ac}$ and indicated by $H_{\parallel}$ as shown in Fig.~\ref{fig:fig2}(a); (ii) along the $z$--axis in the $x$-$z$ plane, perpendicular to $I_{\mathrm{ac}}$ and indicated by $H_{\perp}$. 

The $\Delta\sigma$ of the reference SnTe layer, measured at $T=1.8~K$ for $H_{\parallel}$ and $H_{\perp}$ is reported as a function of the applied $\mu_{0}H$ in Fig.~\ref{fig:fig4}(a). The $\Delta\sigma$ for $H_{\parallel}$ and $H_{\perp}$, are indicated as $\sigma_{H_{\parallel}}$ and $\sigma_{H_{\perp}}$, respectively. As evidenced in Fig.~\ref{fig:fig4}(a), for the $H_{\perp}$ orientation, $\Delta\sigma$ of the reference SnTe layer presents around zero magnetic field an upward cusp: a fingerprint of weak antilocalization (WAL) due to the topological surface states (TSS) stabilized by the $\mathcal{M}$ symmetry in rocksalt SnTe(111) \cite{Taskin:2017_Nat.Commun.,Wang:2018_PRB,Dybko:2017_PRB}. The WAL in the pristine SnTe layer points at the presence of massless DFs in the TSS at the (111) surface. The WAL in topological states of matter, and particularly in SPTs, is an essential feature of the TSS due to self intersecting time reversed paths of the DFs \cite{Hasan:2010_RMP}, characterized by opposite spin states. By fitting the estimated $\Delta\sigma$ according to the theoretical description in terms of the Hikami-Larkin-Nagaoka model \cite{Hikami:1980_PTP}, the prefactor $\alpha$ is estimated to be $\sim0.45$ at $T=1.8~\mathrm{K}$, confirming that the transport in the SnTe layer is due to the TSS. The Berry phase is expressed as:

\begin{equation}
\mathcal{A_\mathrm{b}}=\pi\Big(1-\frac{\Delta}{2E_{\mathrm{F}}}\Big)
\end{equation}

\noindent
with $\Delta$ the energy gap at the Dirac cone and $E_\mathrm{F}$ the Fermi level. Since the TSS residing at the $\left(111\right)$ surface of the SnTe are gapless, then $\mathcal{A_\mathrm{b}}=\pi$ \cite{Bao:2013_Sci.Rep.,Wei:2018_PRB,Nandi:2018_PRB} and the accumulation of a $\pi$ Berry phase leads to the observed WAL. However, by applying an external field $\mu_{0}H$, the destructive interference of the two time reversed paths brings about a negative $\Delta\sigma$, as evidenced in Fig.~\ref{fig:fig4}(a). On the other hand, in the case of the $H_{\parallel}$ orientation, the upward cusp is broadened, pointing at a weakening of the contribution from the 2D massless DFs \cite{Akiyama:2016_Nano,Wang:2018_PRB,Wei:2018_PRB,Ando:2013_JPSJ,Ando:2015_Ann.Rev.}. 

The $\left(111\right)$ surface of SnTe is protected by both $\mathcal{M}$ and $\mathcal{T}$ symmetries. The isothermal $\Delta\sigma$ of the Sn$_{1-x}$Mn$_{x}$Te layers for different $x$ measured at $T = 1.8~\mathrm{K}$ by sweeping $\mu_{0}H$ in the interval $\pm{1}~\mathrm{T}$, is reported in Fig.~\ref{fig:fig4}(b). The introduction of Mn into the SnTe lattice leads on one hand to the breaking of the $\mathcal{T}$ symmetry at the TRIM points of the $\left(111\right)$ surface, while on the other hand it opens a gap at the DP, resulting in $\phi_\mathrm{B}\neq\pi$, and wiping out the WAL. The presence of Mn ions in the SnTe lattice causes the disappearance of the sharp upward cusp near zero field observed for $\Delta\sigma$ in the pristine SnTe layer for both $H_\parallel$ and $H_\perp$. The disappearance of the WAL in Sn$_{1-x}$Mn$_{x}$Te is observed even for $x=0.03$, as seen in Fig.~\ref{fig:fig4}(b). The negative $\Delta\sigma$ of Sn$_{0.97}$Mn$_{0.03}$Te is enhanced by the applied $\mu_{0}H$. The positive $\Delta\sigma$ in degenerate conductors or alloys with magnetic impurities is attributed to weak localization (WL)\cite{Stefanowicz:2014_PRB,Adhikari:2015_PRB}, an orbital mechanism weakening by increasing the applied field, and due to quantum corrections to the conductivity. None of these trends is observed in the case of Sn$_{0.97}$Mn$_{0.03}$Te, where a monotonic increase in the values of negative $\Delta\sigma$ has been observed. The absence of hysteresis in the recorded $\Delta\sigma$ curves indicates that in Sn$_{0.97}$Mn$_{0.03}$Te there is no FM ordering or spin glass phase. The increasing of $x$ in the Sn$_{1-x}$Mn$_{x}$Te lattice results in the positive $\Delta\sigma$ observed for the samples with $x=0.04$, $0.06$, $0.07$ and $0.08$, as reported in Fig.~\ref{fig:fig4}(b). This behavior differs from the one previously reported for magnetically doped TIs like Bi$_{2-x}$Cr$_{x}$Te$_{3}$ \cite{Bao:2013_Sci.Rep.}, where the MC was found to be negative up to Cr concentrations $x\sim0.10$.

The observation of positive $\Delta\sigma$ in Sn$_{1-x}$Mn$_{x}$Te for $x$$\geq$0.04 emphasises the influence of the magnetic impurities Mn on the carrier transport in the presence of $\mu_{0}H$. No hysteretic  behavior of $\Delta\sigma$ has been observed for $x$\,=\,0.04 down to $T$\,= 1.8\,K, as shown in Fig.~\ref{fig:fig4}(b). However, as reported in Fig.~\ref{fig:fig4}(b), an opening of hysteresis in the $\Delta\sigma$ curve is detected for the Sn$_{1-x}$Mn$_{x}$Te layers with $x= 0.06$, $0.07$ and $0.08$, pointing at the onset of FM ordering. The $\Delta\sigma$ for the samples with $x= 0.06$ and $x= 0.07$ as a function of $T$ is shown in Figs.~\ref{fig:fig4}(c) and \ref{fig:fig4}(d), respectively, where an hysteretic behavior is observed for $T\textless5~\mathrm{K}$, suggesting a $T_{\mathrm{c}}\textless5~\mathrm{K}$. The amplitude of the isothermal $\Delta\sigma$ is found to be proportional to $x$, as observed in Fig.\,\ref{fig:fig4}(b). The reduction of carrier scattering from the Mn moments due to an applied $\mu_{0}H$ in a system with magnetic ordering is expected to show such a behavior and it was previously seen in magnetically doped TIs and TCIs \cite{Zhang:2012_PRB,Kilanski:2013_JAP}. With increasing $T$, the magnitude of $\Delta\sigma$ decreases and changes sign from positive to negative for $T\textgreater10~\mathrm{K}$ in the samples with $x$\,=\,0.06 and 0.07, indicating that with increasing $T$ the thermal fluctuations suppress the spin dependent magnetotransport. 

The signature of the onset of a FM ordering and PMA in magnetically doped systems is also manifested through the anisotropic magnetoconductance, when $\Delta\sigma$ is monitored for the two orientations of $\mu_{0}H$, providing $\Delta\sigma_{H_{\parallel}}$ and $\Delta\sigma_{H_{\perp}}$. The $\Delta\sigma_{H_{\parallel}}$ and $\Delta\sigma_{H_{\perp}}$ for the Sn$_{1-x}$Mn$_{x}$Te samples are reported in Fig.~\ref{fig:fig4}(e), where the solid and the dashed lines represent the longitudinal magnetoconductance (LMC) and the transversal magnetoconductance (TMC) respectively, and the applied $\mu_{0}H$ is swept between -1\,T and +7\,T at $T=1.8\,\mathrm{K}$. In the present work, the LMC and TMC are defined as the magnetoconductance measured respectively for $\mu_{0}H\parallel{I_\mathrm{ac}}$ and $\mu_{0}H\perp{I_\mathrm{ac}}$, as reported in Fig.~\ref{fig:fig2}(a).  

%%%%%%%%%%%%%%%%%%%%%%%%%%%%%%%%%%%%%%%%%%%%%%%%%%%%%%%%%%%%%%%%%%%%%
\begin{figure}[htbp]
	\centering
	\includegraphics[scale=2.5]{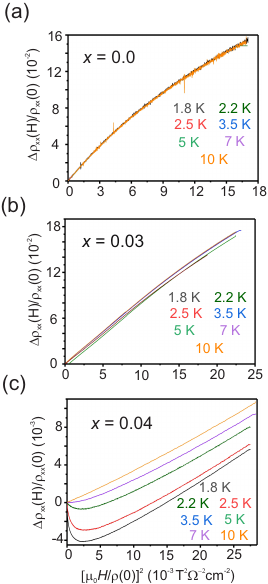}
	%\vspace{3 cm}
	\caption{\label{fig:Kohler} Kohler plots of (a) SnTe, (b) Sn$_{0.97}$Mn$_{0.03}$Te, and (c) Sn$_{0.96}$Mn$_{0.04}$Te epitaxial layers as a function of $T$.}	
	\label{fig:fig5}
\end{figure}
%%%%%%%%%%%%%%%%%%%%%%%%%%%%%%%%%%%%%%%%%%%%%%%%%%%%%%%%%%%%%%%%%%%%%

According to Fig.~\ref{fig:fig4}(e), $\left|\Delta\sigma_{H_{\parallel}}\right|<\left|\Delta\sigma_{H_{\perp}}\right|$ suggests the presence of PMA in the Sn$_{1-x}$Mn$_{x}$Te thin films. It is also observed that $\Delta\sigma$ in samples with $x\geq0.04$ undergoes a transition from positive to negative for $\mu_{0}H$$\geq$1\,T, a behavior assigned to the contributions of both 2D and 3D states, as seen in the case of Mn doped TI systems \cite{Zhang:2012_PRB}. For a strictly spin dependent scattering of the charge carriers in systems with a long range FM order, a transition of $\Delta\sigma$ from positive to negative is not expected \cite{Kilanski:2013_JAP,Zhang:2012_PRB}. The sign of $\Delta\sigma$ for samples with $x=0.06$ and $x=0.07$ is reversed for $\Delta\sigma_{H_{\parallel}}$ in comparison to $\Delta\sigma_{H_{\perp}}$ at $\mu_{0}H$\,=\,7\,T, as reported in Fig.~\ref{fig:fig4}(e). However, for the sample with $x=0.08$, there is no reversal of sign in the measured $\Delta\sigma_{H_{\parallel}}$ and $\Delta\sigma_{H_{\perp}}$.

The change in isothermal longitudinal resistivity $\Delta\rho_\mathrm{xx}$ for a conventional metal in an applied field $\mu_{0}H$ obeys the Kohler's rule \cite{Pippard_Book}, according to the functional relation:

 \begin{equation}
 \frac{\Delta\rho _\mathrm{xx}}{\rho_{0}}=F\Big(\frac{\mu_{0}H}{\rho_{0}}\Big)
  \end{equation}
 
\noindent
where $\rho_{0}$ is the zero-field resistivity at a given $T$. The Kohler's behavior is due to the fact that conventional metals host a single kind of charge carriers. In a weak field limit, the $\Delta\sigma$ of most metals follows a  quadratic dependence on $\mu_{0}H$, \textit{i.e.} $\Delta\sigma\propto\beta{H^2}$, with $\beta$ a proportionality constant. The resistivity is $\rho_{0}\propto\frac{1}{\tau}$, with $\tau$ the scattering time of the itinerant charge carriers in a metallic system such as the one studied here. Therefore, a plot of $\frac{\Delta\rho _\mathrm{xx}}{\rho_{0}}$ \textit{vs.} $\Big(\frac{H}{\rho_{0}}\Big)^{2}$ is expected to collapse into a single $T$-independent curve, provided that the number and type of charge carriers, and the electronic disorder in the system remain constant over the measured $T$ range \cite{Chan:2014_PRL,Spain:1976_Carbon}. The Kohler's plots for SnTe, Sn$_{0.97}$Mn$_{0.03}$Te and Sn$_{0.96}$Mn$_{0.04}$Te are reported in Figs.~\ref{fig:fig5}(a)--\ref{fig:fig5}(c), where it can be seen, that the Kohler's rule is obeyed by SnTe in Fig.~\ref{fig:fig5}(a), while it is violated by Sn$_{0.97}$Mn$_{0.03}$Te and Sn$_{0.96}$Mn$_{0.04}$Te, as observed in Figs.~\ref{fig:fig5}(b) and \ref{fig:fig5}(c), respectively. The deviation from the Kohler behavior is much enhanced in Sn$_{0.96}$Mn$_{0.04}$Te with a significant $T$ dependence of the $\frac{\Delta\rho _\mathrm{xx}}{\rho_{0}}$ \textit{vs.} $\Big(\frac{H}{\rho_{0}}\Big)^{2}$ behavior, pointing at the onset of magnetic ordering in this sample. This inference is corroborated by the fact that  Sn$_{0.96}$Mn$_{0.04}$Te also exhibits a positive $\Delta\sigma$, as reported in Fig.~\ref{fig:fig4}(b). The violation of the Kohler's rule with the introduction of Mn in SnTe suggests, that the scattering times of the charge carriers are modified due to the presence of Mn in the lattice and by the corresponding increment of electronic disorder in the system. It is expected that both the carrier concentration and carrier mobility are affected by the presence of Mn ions in the SnTe lattice. By combining the analysis of $\Delta\sigma$ as a function  of $\mu_{0}H$ and $T$ with the Kohler's plots, it is concluded that a FM order is established in Sn$_{1-x}$Mn$_{x}$Te for $x\geq0.06$ alongside an increased electronic disorder.

\subsection{Anomalous Hall effect}  

%%%%%%%%%%%%%%%%%%%%%%%%%%%%%%%%%%%%%%%%%%%%%%%%%%%%%%%%%%%%%%%%%%%%%
\begin{figure*}[htbp]
	\centering
	\includegraphics[scale=2.0]{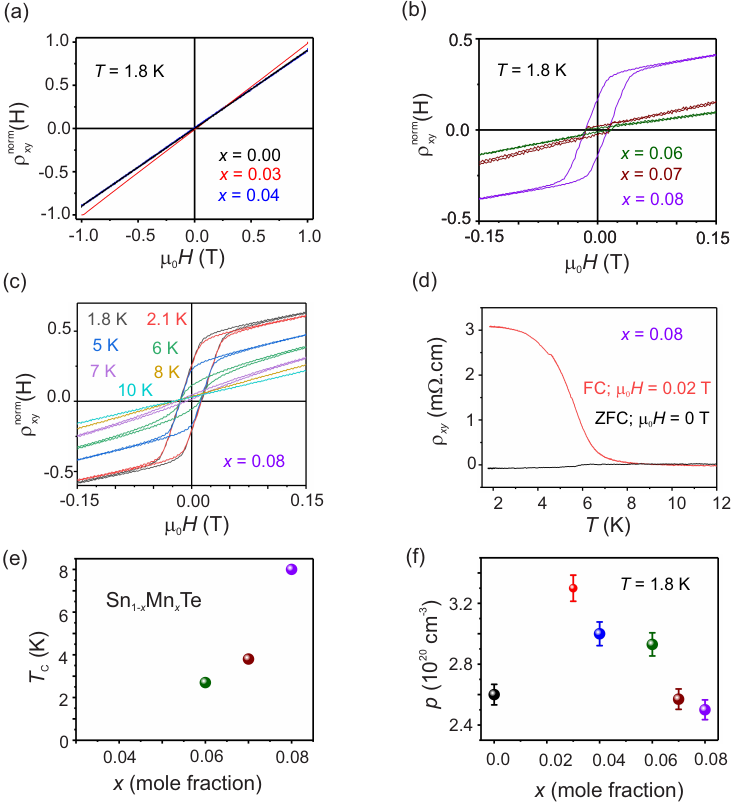}
	%\vspace{3 cm}
	\caption{\label{fig:Hall} Normalised Hall resistivity of Sn$_{1-x}$Mn$_{x}$Te at $T=1.8~\mathrm{K}$ for (a) $x=0.00$, $0.03$, and $0.04$. (b) $x=0.06$, $0.07$, and $0.08$. (c) Normalised Hall resistivity of Sn$_{1-x}$Mn$_{x}$Te for $x=0.08$ as a function of $T$. (d) Field cooled and zero field cooled $\rho_{\mathrm{xy}}$ $vs.$ $T$ for $x = 0.08$ with $\mu_{0}H={0.02~\mathrm{T}}$. (e) Curie temperature --$T_\mathrm{c}$, as a function of Mn content in Sn$_{1-x}$Mn$_{x}$Te samples. (f) Hole concentrations $p_{\mathrm{3D}}$ calculated for the Sn$_{1-x}$Mn$_{x}$Te samples. }	
	\label{fig:fig6}
\end{figure*}
%%%%%%%%%%%%%%%%%%%%%%%%%%%%%%%%%%%%%%%%%%%%%%%%%%%%%%%%%%%%%%%%%%%%%

A spontaneous magnetic order in electronic systems manifests itself through the presence of AHE \cite{Nagaosa:2010_RMP}. In a magnetically ordered regime, the Hall resistivity in a conductor doped with magnetic ions, shows the ordinary Lorentz contribution $R_\mathrm{H}H$ and an additional extraordinary term $R_\mathrm{AH}M$, where $M$ is the magnetization of the system and $R_\mathrm{H}$ and $R_\mathrm{AH}$ are the ordinary and extraordinary Hall or anomalous Hall coefficients, respectively. The AHE in a magnetic conductor originates from asymmetric carrier scattering at magnetic impurity centers when a  magnetic field is applied perpendicular to the applied current. The relation between the $\mathcal{F_\mathrm{b}}$ of the occupied energy bands and the mechanisms of AHE of a magnetic system \cite{Nagaosa:2010_RMP} is essential for the understanding of band topology \cite{Wang:2018_PRB} of SPT phases of matter as the one investigated in this work.

The onset of FM order in a TCI system like SnTe (111) can be probed by studying the signatures of AHE in magnetotransport \cite{Wang:2018_PRB,Ando:2013_JPSJ,Ando:2015_Ann.Rev.,Li:2016_PRL}. The $V_\mathrm{xy}$ of the Sn$_{1-x}$Mn$_{x}$Te thin films has been measured by sweeping $\mu_{0}H$ in the range $\pm{1}$~T. The estimated normalized Hall resistivity $\rho_{xy}^\mathrm{norm}$ at $T=1.8\,\mathrm{K}$ for the Sn$_{1-x}$Mn$_{x}$Te samples with $x = 0.0$, $0.03$ and $0.04$ is shown in Fig.~\ref{fig:fig6}(a) and $\rho_{xy}^\mathrm{norm}$ for $x = 0.06$, $0.07$ and $0.08$ are reported in Fig.~\ref{fig:fig6}(b). The positive slopes of $\rho_{xy}^\mathrm{norm}$ for all the samples under consideration, indicate that the electrical transport in the Sn$_{1-x}$Mn$_{x}$Te system is hole dominated. A $p$-type conduction was already reported for both undoped and magnetically doped bulk and epitaxial thick layers of SnTe \cite{Akiyama:2016_Nano,Wang:2018_PRB,Wei:2018_PRB,Nadolny:2002_JMMM,Dybko:2017_PRB,Eggenkamp:1994_JAP}. A square hysteresis is observed in the evolution of $\rho_{xy}^\mathrm{norm}$ as a function of $\mu_{0}H$, pointing at an hysteretic AHE due to FM ordering in the samples with $x=0.06$, $0.07$ and $0.08$. Unlike proximity induced magnetism at the surface of topological systems, homogeneous doping with magnetic ions of a TCI lattice leads to the simultaneous introduction of gapped Dirac cones at the top and bottom surfaces of the TCI film. Hence, for the Sn$_{1-x}$Mn$_{x}$Te layers, the observation of hysteretic AHE supports the premise of carrier mediated uniform FM ordering in the samples with $x=0.06$, $0.07$ and $0.08$. The Hall resistivities in Fig.~\ref{fig:fig6}(b) are plotted for $-0.15\,\mathrm{T}\leq\mu_{0}H\leq+0.15\,\mathrm{T}$.

%%%%%%%%%%%%%%%%%%%%%%%%%%%%%%%%%%%%%%%%%%%%%%%%%%%%%%%%%%%%%%%%%%%%%
\begin{figure*}[ht]
	\centering
	\includegraphics[scale=1.75]{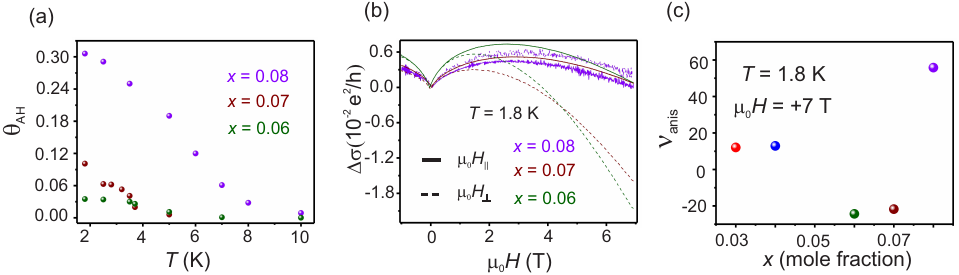}
	%\vspace{3 cm}
	\caption{\label{fig:AHA} (a) Anomalous Hall angle as a function of T for Sn$_{1-x}$Mn$_{x}$Te layers with $x=0.06$, $0.07$ and $0.08$. (b) Magnetic anisotropy of Sn$_{1-x}$Mn$_{x}$Te thin films with $x=0.06$, $x=0.07$ and $x=0.08$ recorded at $T=1.8~\mathrm{K}$. (c) Magnetic anisotropy parameter $\nu_{\mathrm{anis}}$ measured at $T=1.8\mathrm{K}$ and $\mu_{0}H=+7~\mathrm{T}$.}	
	\label{fig:fig7}
\end{figure*}
%%%%%%%%%%%%%%%%%%%%%%%%%%%%%%%%%%%%%%%%%%%%%%%%%%%%%%%%%%%%%%%%%%%%% 

The hysteretic AHE observed for the Sn$_{0.92}$Mn$_{0.08}$Te is comparable to the one reported for Cr doped Bi$_{2}$\big(Se$_{x}$Te$_{1-x}$\big)$_{3}$ \cite{Zhang:2013_Science,Bao:2013_Sci.Rep.}, while weaker hysteresis loops are observed for the Sn$_{0.93}$Mn$_{0.07}$Te and Sn$_{0.94}$Mn$_{0.06}$Te films. A plot of $\rho_{xy}^\mathrm{norm}$ as a function of $\mu_{0}H$ for $x=0.08$, within the range 1.8\,K\,$<$$T$$<$10\,K is provided in Fig.~\ref{fig:fig6}(b). Due to weakening of the  FM ordering caused by thermally assisted spin-fluctuations, a decrease in $\rho_{xy}^\mathrm{norm}$ and in the coercive field $H_{\mathrm{c}}$, is detected for increasing $T$. 

The ferromagnet-paramagnet (FM--PM) phase transition temperature $T_{\mathrm{c}}$ for the Sn$_{1-x}$Mn$_{x}$Te (111) thin films, has been established by measuring $\rho_{\mathrm{xy}}$ as a function of $T$. The samples are cooled down from $T$\,=20\,K to $\sim$1.8\,K with a temperature ramp of $0.5~\mathrm{K}/\mathrm{min}$, both in the presence and in the absence of $\mu_{0}H$. The field cooling (FC) is carried out at $H_{\perp}=0.02~\mathrm{T}$. The $\rho_\mathrm{xy}$\,$vs.$\,$T$ trend for $x=0.08$ is reported in Fig.~\ref{fig:fig6}(c), revealing a Brillouin-like behavior with  $T_{\mathrm{c}}\sim{7.5}~\mathrm{K}$.  The measured $T_\mathrm{c}$ as a function of $x$ is provided in Fig.~\ref{fig:fig6}(e). The highest $T_\mathrm{c} = 7.5~\mathrm{K}$ is found for the sample with $x=0.08$, while the transition temperatures for $x=0.06$ and $x=0.07$ are $T_\mathrm{c}$\,=\,3\,K and 3.5\,K, respectively. No fingerprint of a FM--PM phase transition has been observed for the samples with $x=0.03$ and $x=0.04$. Conclusions on either a SG or on a RSG phase transition cannot be driven from the $\rho_\mathrm{xy}$\,$vs.$\,$T$ analysis.

For $T<T_\mathrm{c}$, the contribution to the total Hall effect is dominated by AHE. For a 6-contact Hall bar, $\rho_{\mathrm{xy}}$ is expressed as:

\begin{equation}
\rho_{\mathrm{xy}}\left(H\right)=R_\mathrm{H}H+\mu_{0}R_\mathrm{AH}M
\end{equation}

\noindent 
The bulk hole concentration $p_{3\mathrm{D}}$, calculated from the measured Hall coefficients $R_{\mathrm{H}}$ and $R_\mathrm{AH}$ as a function of $x$ at $T=1.8~\mathrm{K}$ is provided in Fig.~\ref{fig:fig6}(f). The estimation of $p_{3\mathrm{D}}$ is carried out for $-1\,\mathrm{T}\leq\mu_{0}H\leq+1\,\mathrm{T}$ --where $\rho_{\mathrm{xy}}$ is linear-- and $p_{3\mathrm{D}}$ at $T$\,=\,1.8\,K for the 30\,nm SnTe (111) film is $\sim$$\left(2.6\times10^{20}\right)$~cm$^{-3}$, in agreement with similar reports on the hole concentration in bulk \cite{Chi:2016_APL} and $\sim$2\,$\mu$m thick SnTe epitaxial layers \cite{Nadolny:2002_JMMM,Dybko:2017_PRB,Nadolny:2002_JMMM}. The substantial $p_{3\mathrm{D}}$ in SnTe is due to the presence of Sn vacancies in the lattice. A slight increase of $p_{3\mathrm{D}}$ to $\sim$(3.3$\times$10$^{20}$)\,cm$^{-3}$ for $x$\,=\,0.03 is observed. However, for $x\textgreater0.03$ in Sn$_{1-x}$Mn$_{x}$Te, the values of $p_{3\mathrm{D}}$ decrease monotonically as a function of $x$. Both an increase \cite{He:2015_JMCA} or a decrease of $p_{3\mathrm{D}}$ \cite{Chi:2016_APL} with $x$ have been reported for bulk Sn$_{1-x}$Mn$_{x}$Te, depending on the density of Sn vacancies. In the case of SnTe epitaxial thin films, by tuning the Te flux the density of Sn vacancies can be adjusted to achieve the required hole concentrations \cite{Nadolny:2002_JMMM}.  
During the epitaxial process for the samples studied here, the flux from the elemental Te source in combination with the one from the SnTe compound source has allowed a controlled adjustment of the stoichiometry and of the density of Sn vacancies, which in turn leads to the observed high value of $p_{3\mathrm{D}}$ in the SnTe reference. By tuning the Te flux also during the growth of Sn$_{1-x}$Mn$_{x}$Te, a nominal density of vacancies comparable with the one in SnTe is ensured. However, the substitutional inclusion of Mn leads to a decrease in the density of Sn vacancies, causing a lowering of $p_{3\mathrm{D}}$ with increasing $x$. 

The magnetic phase diagram of Sn$_{1-x}$Mn$_{x}$Te has been reported to exhibit both long and short range FM ordering, spin glass and re-entrant spin glass phases \cite{Eggenkamp:1995_PRB}, with the exchange mechanism largely determined by an interplay of $x$ and $p_{3\mathrm{D}}$.  Theoretical calculations predict short range FM order for samples with $p_{3\mathrm{D}}$\,$\sim$\,(2$\times$10$^{20}$)\,cm$^{-3}$ and Mn doping, $x$$\sim$(0.03-0.10), while for $p_{3\mathrm{D}}$$\sim$10$^{21}$\,cm$^{-3}$ a re-entrant spin glass phase is expected \cite{Eggenkamp:1995_PRB,Story:1993_PRB}. For the samples studied in this work, the estimated $p_{3\mathrm{D}}$$\sim$(3$\times$10$^{20}$) and $x$$\sim$(0.03-0.08), and therefore the observed FM ordering of the Sn$_{1-x}$Mn$_{x}$Te thin films is attributed to hole-mediated RKKY interaction, similarly to the case of bulk and thick films \cite{Zhu:2011_PRL, Wang:2018_PRB,Kou:2015_SSC,Eggenkamp:1995_PRB,Story:1993_PRB}. It is also worth noting, that $T_{\mathrm{c}}$ increases with $x$, even though a decrease of $p_{3\mathrm{D}}$ is observed for increasing $x$.

The ferromagnetic Hall response of an itinerant ferromagnetic system is expressed in terms of the anomalous Hall angle $\theta_{\mathrm{AH}}$ \cite{Nagaosa:2010_RMP,Kim:2018_Nat.Mater} defined as:

\begin{equation}
\theta_{\mathrm{AH}}=\frac{\sigma_{\mathrm{xy}}^{\mathrm{AH}}}{\sigma_{\mathrm{xx}}}
\end{equation}

\noindent
where $\sigma_{xy}^{\mathrm{AH}}$ is the anomalous Hall conductance and $\sigma_{\mathrm{xx}}$ the longitudinal conductance. Thus, $\theta_{\mathrm{AH}}$ is a measure of the anomalous Hall current with respect to the conventional current. A substantial value of $\theta_{\mathrm{AH}}$ indicates the immunity of the system to impurity scattering \cite{Kim:2018_Nat.Mater,Singha:2019_PRB,Nagaosa:2010_RMP}. The calculated values of $\theta_{\mathrm{AH}}$ for the ferromagnetic Sn$_{1-x}$Mn$_{x}$Te layers with $x=0.06$, $0.07$ and $0.08$ as a function of $T$ at $\mu_{0}H=1\,\mathrm{T}$ are reported in Fig.~\ref{fig:fig7}(a) and the sample with the highest $x$ has at $T$\,=1.8\,K the greatest value $\theta_{\mathrm{AH}}$$\sim$0.3, higher than the ones reported for the topological ferromagnetic semimetal Fe$_{3}$GeTe$_{2}$ \cite{Kim:2018_Nat.Mater} and for the half Heusler ferromagnet TbPtBi \cite{Singha:2019_PRB}. The reported $\theta_{\mathrm{AH}}$ for both Fe$_{3}$GeTe$_{2}$ \cite{Kim:2018_Nat.Mater} and TbPtBi \cite{Singha:2019_PRB} were measured at $\mu_{0}H$$\gg$1\,T, whereas in the case of the Sn$_{1-x}$Mn$_{x}$Te layers considered here, the estimation of $\theta_{\mathrm{AH}}$ is carried out for $\mu_{0}H$\,=\,1\,T. The large value of $\theta_{\mathrm{AH}}$ for $x=0.08$ indicates that the system is insensitive to impurity scattering, in spite of the electronic disorder defined by $\left(\frac{1}{\mathrm{RRR}}\right)$ w.r.t. $x$ and shown in Fig.~\ref{fig:fig3}(a). 
An anisotropy parameter, $\nu_{\mathrm{anis}}=\frac{\Delta\sigma_{H_{\perp}}}{\Delta\sigma_{H_{\parallel}}}$ which essentially represents the ratio of the measured TMC to the LMC is defined to quantify the magnetic anisotropy in the studied samples. The parameter $\nu_{\mathrm{anis}}$ is estimated as a function of $x$ at $\mu_{0}H$\,=\,+7\,T and $T$\,=\,1.8\,K from the $\Delta\sigma$ \textit{vs.} $\mu_{0}H$ plots shown in Fig.~\ref{fig:fig7}(b) for the samples exhibiting FM order. The negative values of $\Big[\nu_{\mathrm{anis}}\Big]_{H=7\mathrm{T}}$ for $x=0.06$ and for $x=0.07$ reported in Fig.\,\ref{fig:fig7}(c) relate to the sign reversal of $\Delta\sigma$ from positive to negative for the in-plane and out-of-plane $\mu_{0}H$ orientations. For the sample with $x$\,=\,0.08, $\nu_{\mathrm{anis}}$ is calculated to be +55 and the sample shows a positive $\Delta\sigma$ even at $\mu_{0}H$$\gg$1\,T. The values of $\Delta\sigma_{H_{\perp}}$ and $\Delta\sigma_{H_{\parallel}}$ are inverted in the sample with $x=0.08$ pointing at a dominant out-of-plane magnetic contribution. Therefore, longitudinal and transverse magnetotransport measurements on 30 nm thin Sn$_{1-x}$Mn$_{x}$Te $\left(111\right)$ films with $x>0.06$ give quantitative information on $\Delta\sigma$, $\theta_{\mathrm{AH}}$, presence of PMA and electronic disorder, and point to a topology driven ferromagnetic phase transition in the studied magnetically doped TCI systems. The obtained parameters satisfy the necessary criteria for epitaxial  Sn$_{1-x}$Mn$_{x}$Te $\left(111\right)$ to host topologically protected QAH states.

\section{Conclusion} 

In summary, high quality epitaxial thin layers of Sn$_{1-x}$Mn$_{x}$Te (111) have been grown in an MBE system and low $T$-high $\mu_{0}H$ magnetotransport studies on lithographically fabricated Hall bar devices are conducted. All the samples considered in this work show metallic behavior with hole concentrations $p_{3\mathrm{D}}$$\sim$(2$\times$10$^{20}$)\,cm$^{-3}$. The decrease of the inverse of residual resistivity ratio parameter $\left(\frac{1}{\mathrm{RRR}}\right)$ as a function of $x$, is an indication of the enhanced disorder with increasing $x$ in Sn$_{1-x}$Mn$_{x}$Te. With Mn doping of SnTe, a gradual change of the sign of $\Delta\sigma$ from negative to positive points at the onset of spin mediated magnetoconductance. For $x\geq0.04$, a hysteretic butterfly-like behavior of $\Delta\sigma$ is a signature of the emergence of FM ordering in the system. The AHE with a square hysteresis below $T_{\mathrm{c}}$ measured for the samples with $x\ge{0.06}$ lets infere an ordered FM interaction in the samples.  A maximum $T_{\mathrm{c}}\sim~7.5~\mathrm{K}$ is established for $x=0.08$, while for $x=0.06$ and $x=0.07$, $T_{\mathrm{c}}\textless3.5~\mathrm{K}$ have been estimated by measuring $\rho_{\mathrm{xy}}$ as a function of $T$. The value of $\theta_{\mathrm{AH}}=0.3$ for the sample with $x=0.08$ is higher than the ones found for the samples with $x=0.07$ and $x=0.06$, in which $\theta_{\mathrm{AH}}=0.1$ and $\theta_{\mathrm{AH}}=0.03$ respectively. The remarkable anomalous Hall angle found for the topology driven ferromagnetic TCI Sn$_{0.92}$Mn$_{0.08}$Te thin film is among the highest reported for magnetic topological states of matter. Hence, it is concluded that the magnetic Sn$_{1-x}$Mn$_{x}$Te layers with $x\geq0.06$ with disorder immune anomalous Hall transport can realize topologically protected QAH states. Therefore, while designing a magnetically doped TCI system to stabilize a QAH state, an optimization of the critical parameters including high $T_{\mathrm{c}}$, $p_{3\mathrm{D}}$, $\theta_{\mathrm{AH}}$ and $\nu_{\mathrm{anis}}$ is mandatory. By systematically tuning the chemical potential of Sn$_{1-x}$Mn$_{x}$Te either \textit{via} doping with Pb or Bi or by employing electrostatic gating \cite{Wang:2018_PRB,Ando:2015_Ann.Rev.}, a high--$\mathcal{C}$ QAH state with $\mathcal{C}>1$ can be realized in epitaxial Sn$_{1-x}$Mn$_{x}$Te layers with FM order. 
 
\section*{Acknowledgements}

This work was funded by the Austrian Science Fund (FWF) through Projects No. P26830 and No. P31423. The technical support by Ursula Kainz (optical lithography and patterning) and Philip Lindner (all issues related to the magneto-transport system) is gratefully acknowledged. The research of V.V. Volobuev was supported by the Foundation for Polish Science through the IRA Programme co-financed by the EC within SGOP.

%\bibliographystyle{abbrv}  
%\bibliographystyle{apsrev4-2} 

%\bibliographystyle{apsrev4-1} 
%\bibliography{SnMnTe_PRX_bib}

\end{document}